\documentclass[a4paper]{jpconf}
\usepackage{graphicx}
\begin{document}
\title{Evidence of Magnetic Correlations Among the Charge Stripe Electrons of Charge-Stripe Ordered La$_{2}$NiO$_{4.11}$. }

\author{P. G. Freeman$^{1,2}$, S. M. Hayden$^{3}$, R. A. Mole$^{4,5}$, M. Enderle$^{2}$, D. Prabhakaran$^6$and A. T. Boothroyd$^6$. }


\address{$^1$Helmholtz-Zentrun Berlin f\"{u}r Materialien und Energie GmbH, Hahn-Meitner-Platz 1, D-14109 Berlin, Germany}

\address{$^2$Institut Laue-Langevin, BP 156, 38042 Grenoble Cedex 9, France}

\address{$^3$Department of Physics, University of Bristol, Bristol, BS8 1TL, United Kingdom}

\address{$^4$Bragg Institute, ANSTO, New Illawarra Road, Lucas Heights, NSW,  Australia}

\address{$^5$FRM II, Lichtenbergstra\ss e, 1 85747 Garching, Germany}

\address{$^6$Department of Physics, Oxford University, Oxford, OX1 3PU, United Kingdom}

\ead{paul.freeman@helmholtz-berlin.de}

\begin{abstract}

The magnetic excitations of charge-stripe ordered La$_{2}$NiO$_{4.11}$ where investigated using polarized- and unpolarized-neutron scattering to determine the magnetic excitations of the charge stripe electrons. We observed a magnetic excitation mode consistent with the gapped quasi-one-dimensional antiferromagnetic correlations of the charge stripe electrons previously observed in La$_{2-x}$Sr$_{x}$NiO$_{4}$ $x = 1/3$ and $x = 0.275$.


\end{abstract}

\section{Introduction}

Recent observations of a 
hourglass shaped excitation spectrum in cuprate superconductors point towards an underlying universal magnetic excitation spectrum for all of the cuprates\cite{hourglass}. In insulating charge stripe ordered La$_{5/3}$Sr$_{1/3}$CoO$_{4}$ an hourglass magnetic excitation spectrum has been observed, and explained within a charge stripe model with a weak inter-stripe spin interaction, relative to the intra-stripe spin interaction\cite{boothroyd-Nature}. The magnetic excitations of La$_{5/3}$Sr$_{1/3}$CoO$_{4}$ indicate that the underlying magnetism of the cuprates can originate from charge stripe fluctuations. Furthermore, the lack of an hour glass excitation spectrum in charge stripe ordered La$_{2-x}$Sr$_{x}$NiO$_{4}$ is explained by the strong inter-stripe spin interaction of these materials. 

The dispersion of the magnetic excitations of charge stripe ordered  La$_{2}$NiO$_{4.11}$  deviates from the theory  that describes the magnetic excitations in La$_{5/3}$Sr$_{1/3}$CoO$_{4}$ and La$_{2-x}$Sr$_{x}$NiO$_{4}$ $x  \sim 1/3$ \cite{freeman-PRB-2009}. In a limited energy range, $\sim$10-30\,meV, the centre of the magnetic excitations shifts inwards towards the two dimensional antiferromagnetic wavevector ${\bf Q_{AFM}} = (0.5, 0.5)$. If the inter-stripe spin interaction was weak, the excitations would be expected to disperse towards  ${\bf Q_{AFM}}$, but the inter-stripe spin interaction of La$_{2}$NiO$_{4.11}$  is strong\cite{freeman-PRB-2009,Yao-PRL-06}. As the inter-stripe spin interaction is mediated by the charge stripe electrons, it is important to ascertain what are the magnetic properties of the charge stripe electrons of La$_{2}$NiO$_{4.11}$. In  La$_{2-x}$Sr$_{x}$NiO$_{4}$ $x \sim 1/3$ we have observed the charge stripe electrons have gapped  quasi-one-dimensional antiferromagnetic spin correlations (q-1D) along the charge stripes\cite{boothroyd-PRL-2003,freeman-PRB-2011}.


\section{Experimental details}

A single crystals of La$_{2}$NiO$_{4.11}$ ($\delta = 0.11$) was
grown by the floating-zone technique\cite{prab}. The sample used
here was  also used in our previous neutron studies of $\delta = 0.11$
\cite{freeman-PRB-2009}, and is a rod of 8\,mm
diameter and 40\,mm length. Thermo-Gravimetric Analysis (TGA) of an
as-grown crystal determined the oxygen excess to be $\delta = 0.11\,
\pm\,0.01$. Data on the bulk magnetization of an as-grown crystal
are published elsewhere\cite{freeman-PRB-2006}. 

Neutron scattering measurements were performed on the triple-axis spectrometers  IN20 at the 
ILL, and on PUMA at FRM-II. The energies of the incident and scattered
neutrons were selected by Bragg reflection from crystal Heusler arrays (IN20), or  arrays of
pyrolytic graphite crystals (PUMA).  The monochromators were horizontally focused on both instruments, and vertically  focused on PUMA,  to maximise neutron flux on the sample position.
The analyzers were horizontally focused on both instruments and
vertically focused on PUMA.
Data were collected with a fixed final neutron wavevector of
2.662\,\AA$^{-1}$. A pyrolytic graphite filter was placed between
the sample and analyzer to suppress higher-order harmonic
scattering. On IN20 polarized neutrons were employed, and the
neutron spin polarization ${\bf P}$ was maintained in a specified
orientation with respect to the neutron scattering vector ${\bf Q}$
by an adjustable guide field of a few mT at the sample position. A monitor is placed between the monochromator and sample position to determine the number of  neutrons incident on the sample position. 
We aligned the crystal so that the horizontal scattering plane was
$(h,h,l)$ (we refer here to the tetragonal unit cell of the space group $I4/mmm$ with unit
cell parameters $a=3.8$\,{\AA} and $c=12.7$\,{\AA}). On IN20 the sample was mounted in a standard orange cryostat, and on PUMA the sample was mounted in a closed cycle refrigerator.

\section{Results}


We performed longitudinal polarized neutron scattering measurements on IN20. For these measurements we constrained the neutron polarization ${\bf P}$ to be parallel to the scattering
vector ${\bf Q}$. In this configuration scattering from electronic magnetic moments causes the neutron spin to flip, whereas scattering via non-magnetic processes does not.

\begin{figure}[!h]
\begin{center}
\includegraphics[clip=,width = 8 cm]{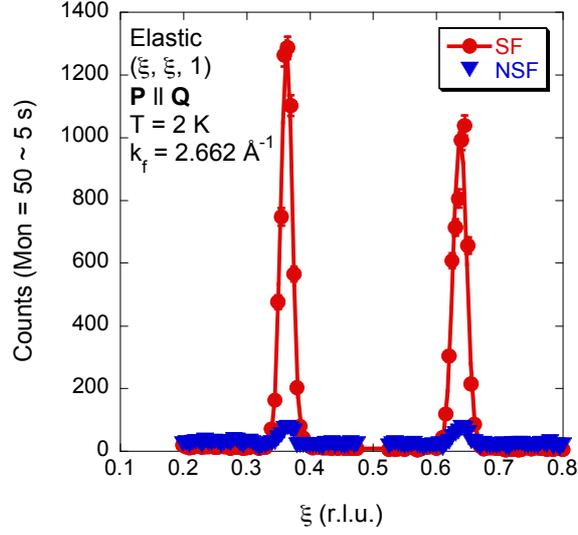}
\caption[Elastic polarised neutron scan of La2NiO4 ]{Spin-flip (SF) and non-spin-flip (NSF)
diffraction from La$_{2}$NiO$_{4.11}$ at T = 2 K. The scan is
parallel to $(hh0)$ with $l = 1$. No correction has been made for
the imperfect polarization of the neutron beam. The solid line is the result of a fit to the SF data with 2 Gaussian lineshapes on a flat background.} \label{fig:elastic}
\end{center} \end{figure}

In figure \ref{fig:elastic}  we show an elastic scan of the $\delta = 0.11$  parallel to $(hh0)$ through the the magnetic Bragg reflections at $l = 1$. In the Spin Flip (SF) data two sharp Bragg reflections are observed at $\xi = 0.363 \pm 0.001$\, r.l.u. and $\xi = 0.637 \pm 0.001$\, r.l.u.  from the magnetic order. In the Non-Spin Flip channel (NSF) weaker Bragg reflections are observed due to the imperfect polarization of the neutron beam. The ratio of the intensities of the NSF and SF Bragg reflections gives a flipping ratio of 19.

\begin{figure}[!h]
\begin{center}
\includegraphics[clip=, width = 15cm]{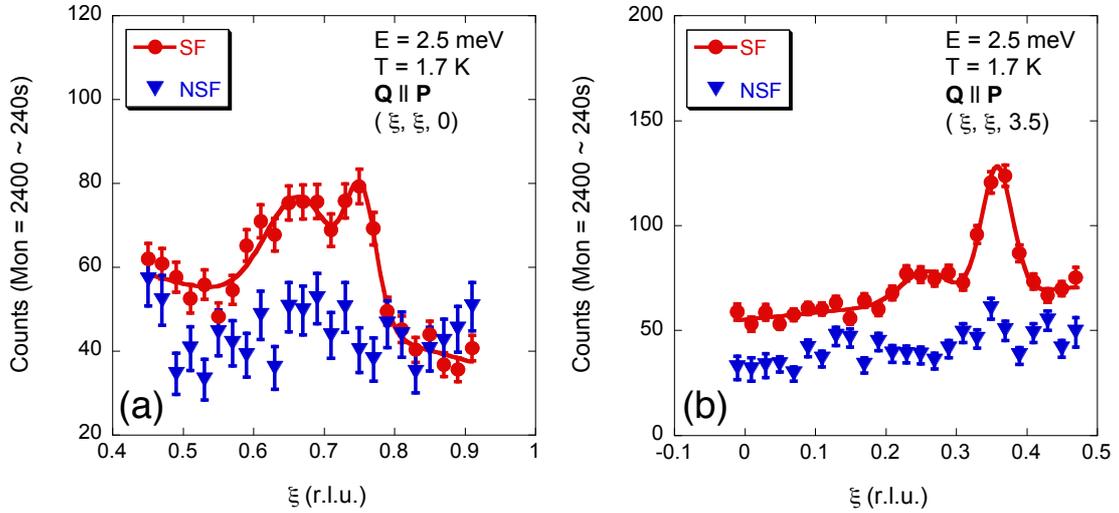}
\caption[Inelastic polarised neutron scans of La2NiO4 ]{Spin-flip (SF) and non-spin-flip (NSF) scattering of $\delta = 0.11$ at 2.5\, meV along directions parallel to $(h,h,0)$ for (a) $l = 0$, and (b) $l = 3.5$. The solid line  is a fit to the SF data of 2 Gaussian lineshapes on a sloping background. The sense of the two peaks is reversed in the two scans, i.e. the ordered Ni$^{2+}$ spins result in the left excitation peak in (a), and the right peak in (b). 
} \label{fig:polinelastic}
\end{center} \end{figure}

In figure \ref{fig:polinelastic}(a) we show scans along $(hh0)$ at $E = 2.5$\, meV   of  $\delta = 0.11$, 
for both the SF and NSF channels.  The data has been corrected for  the imperfact  neutron polariztion\cite{fr}. 
In the NSF channel no excitation peaks are observed, whereas in the SF channel a broad magnetic excitation peak centred on $\sim 0.68$\, r.l.u.  is observed. 
As the dispersion gradient of excitations from the ordered Ni$^{2+}$ $S = 1$ ordered spins is effectively infinite at 2.5\,meV, they only account for the left hand side of the  SF peak. The right hand side of the excitation  centred on $\xi \sim 0.75$ is an unaccounted for antiferromagnetic excitation, that has no elastic component.


\begin{figure}[!hb]
\begin{center}
\includegraphics[clip=, width = 15cm]{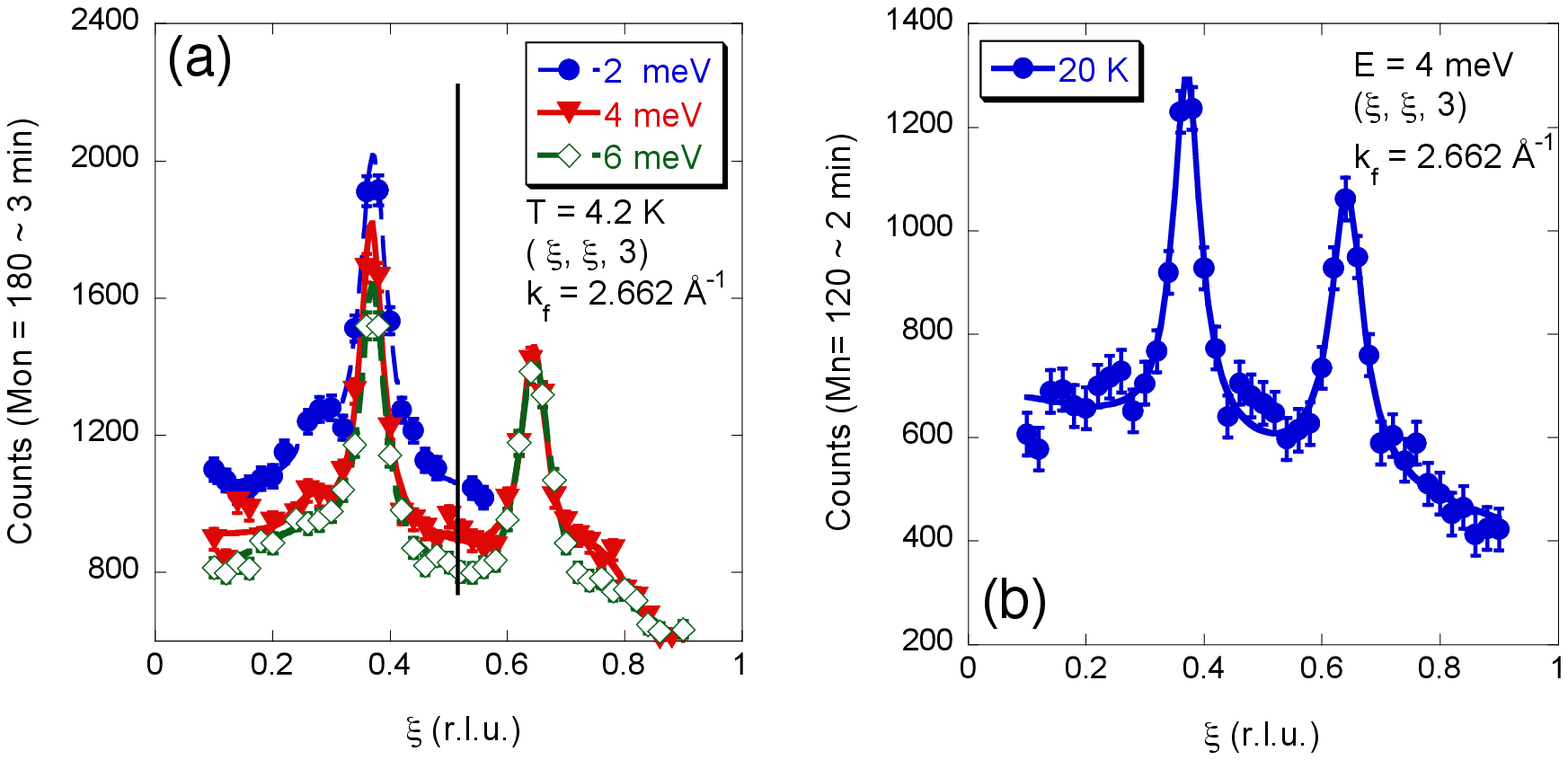}
\caption[Theenergy and temperature dependence of low energy magnetic excitations in La2NiO4.11]{(a) Scans parallel to $(h,h,0)$ of the magnetic excitations of $\delta = 0.11$ at T= 4.2\, K.
A solid vertical line indicates the left and right hand sides of the scans, that were fitted independently.
The other solid and dashed lines indicate fits to the data of 2 Lorentzian lineshapes on a sloping background. (b) A scan parallel to $(h,h,0)$ of the magnetic excitations of $\delta = 0.11$ for 4.0\,meV taken at 20\,K. The solid line is a fit to the data with 2 Lorentzian lineshapes on a sloping background.}
\label{fig:EandT}
\end{center} \end{figure}

At 2.5\,meV we are below the out of plane anisotropy gap in $\delta = 0.11$, and the spins fluctuate around the ordered moment in the Ni-O plane\cite{freeman-PRB-2009}. For a wavevector along $(hh0)$  at 2.5\,meV only the component of the spin fluctutaions perpendicular to this direction are probed. If the spin excitations are observed at a wavector of the same length that is appproximately along  $(00l)$, all the in plane spin fluctuations are probed and the intensity of the observed excitation  will increase. The q-1d in the $x = 1/3$ was determined to preferentially fluctuate out of the Ni-O plane\cite{boothroyd-PRL-2003}. If the q-1D is measured at a wavector appproximately along  $(00l)$ the intensity will decrease compared to a wavevector along $(h,h,0)$. Figure \ref{fig:polinelastic} (b) shows a polarized neutron scan of the magnetic excitations along $(\xi,\ \xi,\ 3.5)$ which shows two peaks in the SF channel. In Fig.\ref{fig:polinelastic} (a) and (b) the symmetry of scans reverses, so that the right hand side peak in (b) is from the ordered Ni$^{2+}$ spins, which has increased in intensity as expected. If the SF peak on the left of Fig.\ref{fig:polinelastic} (b) was from q-1D excitations that preferential fluctuate out of the Ni-O plane, the intensity of these excitations will be weaker  in Fig.\ref{fig:polinelastic} (b) than Fig.\ref{fig:polinelastic} (a), as observed. 

The polarized neutron scattering  measurements of  $\delta = 0.11$ at 2.5\,meV
show two magnetic excitations. We have shown that one of these excitation modes is gapped with no elastic component, and has the same wavevector dependence  for intensity as the q-1D in $x = 1/3$ and $ x = 0.275$. After confirming the existence of a second magnetic excitation in $\delta = 0.11$, we studied the energy and temperature dependence of this magnetic excitation using unpolarized neutron scattering.



Figure \ref{fig:EandT} (a) we show three scans parallel to $(h,h,0)$ at $l = 3$ of the $\delta = 0.11$ at 4.2\,K. As we were able to collect significant neutron counts in the data, we clearly observe the Lorentzian lineshape of the excitations, consistent with our previous results\cite{freeman-PRB-2009}.  At 2\,meV and $\xi = 0.270 \pm 0.006$ a small side peak is clearly resolved from the excitations from the order moments. The centring of this excitation agrees with the flexing of the q-1D observed in $x \sim 1/3$\cite{boothroyd-PRL-2003,freeman-PRB-2011}. At 4\,meV the excitations at  $\xi = 0.270$ and  $\xi = 0.73$ are observed to have reduced counts compared to 2\,meV, and by 6\,meV the excitations have broadened to become an extended tail to the excitation peaks from the ordered moments. When the q-1d of $x = 1/3$ is  scanned in an equivalent way to Fig. \ref{fig:EandT} (a), the q-1d is qualitatively observed to follow the same trend. Unfortunately the fitting uncertainties for the data fits  of the 4\,meV and 6\,meV data in Fig. \ref{fig:EandT} (a) do not allow for a meaningful quantative comparison.

In figure \ref{fig:EandT} (b) we show a scan parallel to $(h,h,0)$ at $l = 3$ for 4\,meV of the $\delta = 0.11$, taken at 20\,K. The scan at 20\,K is well described by a fit of two Lorentzians on a sloping background, as indicated by the solid line of Fig. \ref{fig:EandT} (b). In the $x = 1/3$, on increasing temperature  the q=1D remains observable to temperatures well in excess of 20\,K\cite{freeman-JPCM-2008}, unlike the scan at 20\,K of the $\delta = 0.11$.  
The  lack of observable q-1D at 20\,K for $\delta = 0.11$ may be a result of the fit, with the sloping background and the Lorentzian lineshape incorrectly accounting for the q-1D. Additional measurements in the $(hk0)$ scattering plane away from the excitations from the order Ni$^{2+}$ spins, should resolve this issue. 

\section{Discussion and Conclusions}

We have presented experimental evidence of a second magnetic excitation mode in charge stripe ordered  La$_{2}$NiO$_{4.11}$. A preferential polarization of the fluctuations  out of the Ni-O plane, and the energy dependence of the  magnetic excitations is consistent with the gapped quasi-one-dimensional antiferromagnetic excitations of the charge stripe electrons (q-1D)  in La$_{2-x}$Sr$_{x}$NiO$_{4}$ $x \sim 1/3$. The main components of the ground state magnetic excitations in  La$_{2}$NiO$_{4.11}$,  the out of plane anistropy gap, the inter- and intra-stripe interactions, and the q-1D observed in this work,  are unchanged compared to La$_{2-x}$Sr$_{x}$NiO$_{4}$ $x \sim 1/3$\cite{freeman-PRB-2009,boothroyd-PRL-2003,freeman-PRB-2011}.

Despite the lack of differences in the  magnetic interactions of La$_{2}$NiO$_{4.11}$  and La$_{2-x}$Sr$_{x}$NiO$_{4}$ $x \sim 1/3$, the ground state magnetic excitations from the order Ni$^{2+}$ spins  in the energy range 10-30\,meV have remarkable differences. In La$_{2}$NiO$_{4.11}$ there is an inward dispersion of the magnetic excitations towards ${\bf Q_{AFM}} = (0.5, 0.5)$\cite{freeman-PRB-2009}, whereas in La$_{2-x}$Sr$_{x}$NiO$_{4}$ $x \sim 1/3$ there is a partial gapping of the spin excitations accompanied by a resonance like feature at the magnetic zone centre\cite{freeman-JPCM-2008,boothroyd-PRB-2003,boothroyd-PhysicaB,Woo}. As the main magnetic interactions in these two materials are unaltered, we require further interactions to fully explain the magnetic excitation spectrums in the charge stripe ordered nickelates. As our previous studies for $x \sim 1/3$ have indicated, we need to research beyond the magnetic interactions to understand these striking phenomena  \cite{freeman-JPCM-2008,boothroyd-PRB-2003,boothroyd-PhysicaB}.

\section{Acknowledgements}

This research project has been supported by the European Commission under the 7th Framework Programme through the 'Research Infrastructures' action of the 'Capacities' Programme, Contract No: CP-CSA\_INFRA-2008-1.1.1 Number 226507-NMI3. Crystal growth was supported by the Engineering and Physical Sciences Research Council of Great Britain.


\begin{thebibliography}{10}
\bibitem{fr} 
A flipping ratio of f = 15 was used to correct the data, realistically reduced from the value of 19 observed in diffraction. Correcting the data using $f= 10-19$ results in insignificant differences to the data of Fig. \ref{fig:polinelastic}.
\end{thebibliography}
\end{document}